\documentclass[a4paper, 12pt]{article}
\usepackage[margin=3cm]{geometry}

\usepackage{graphicx}
\usepackage{lipsum}
\usepackage{amsfonts,bbm,amssymb}
\usepackage{makeidx}

\title{Testing the Influence of the Gravitational Redshift on the Casimir Effect in Space}
\author{Clovis Jacinto de Matos, Martin Tajmar \\ Institute of Aerospace Engineering \\ Technische Universit\"at Dresden, Germany}

\begin{document}

\maketitle

\begin{abstract}
We show that the Casimir effect should be influenced by variations of the gravitational potential. This could be tested with a satellite in a highly elliptic orbit. Still significant technology development is required to achieve a relative accuracy of $<5 \times 10^{-10}$ necessary to detect the effect around Earth. That may be obtained in the future based on recent concepts for giant casimir forces. Such a space mission could evaluate both the laws of gravitation and quantum mechanics and their interaction. A dedicated mission, Gravity-Probe C, is proposed.
\end{abstract}

\maketitle

\section*{Introduction}
Two parallel conducting plates separated by a distance $a$ constrain the possible fluctuations of the electromagnetic quantum vacuum to certain stationary resonating values, thus creating a pressure gradient of zero-point energy between the interior and the exterior regions of the cavity. This translates into an attractive force $F$ between the two conducting plates which is directly proportional to their Area $A$ and inversely proportional to the fourth power of their separating distance.
\begin{equation}
F=\frac{\pi^2}{240}\frac{\hbar c}{a^4} A\label{e1}
\end{equation}
where $\hbar$ and $c$ are respectively the Planck Constant and the speed of light in vacuum. This effect was first predicted by H. Casimir in 1948 \cite{Casimir} and has been experimentally confirmed since the 1990s \cite{Bordag2009}. It is currently considered as one of the most fundamental experimental demonstrations of the quantum nature of vacuum.

In the theory of General Relativity, Einstein associates the vacuum to the geometry of spacetime, which also accounts for gravitation through the equivalence principle. Thus in a certain sense vacuum is gravitation. This together with the principle of energy conservation causes the distortion of the wavelength $\lambda$ of an electromagnetic wave as it propagates in a varying gravitational potential $\Phi=-GM/r$ produced by a body with mass $M$. The gravitational distortion of the light wavelength $\lambda=cT$ can be interpreted either as a change in the velocity of light at the place where the gravitational potential is $\Phi$, or as a change in the light period, i.e. as a frequency shift caused by the gravitational potential \cite{Einstein}. Both choices are correct and fully transparent with respect to each other\cite{Born}. Both interpretations reflect the dispersive nature of gravitation when distorting the Minknowski vacuum (which in the absence of gravitation is simply flat spacetime) \cite{Minkowski}. The Gravitational frequency shift reflects the distortion of proper time by the gravitational potential, it is usually designated as the gravitational redshift,
\begin{equation}
\frac{\delta \nu}{\nu_0}=\frac{\Phi}{c^2}\label{e2}
\end{equation}
where $\delta \nu = \nu - \nu_0$,  $\nu_0$ is the frequency at the origin of coordinates where the gravitational potential is gauged to zero, $\nu$ is the frequency at the place where one observes a gravitational potential $\Phi$. A varying speed of light (changing refractive index of the vacuum) is observed through the bending of light near a gravitating mass.
\begin{equation}
\frac{\delta c}{c_0}=\frac{\Phi}{c_0^2}\label{e3}
\end{equation}
where $\delta c = c - c_0$,  $c_0$ is the speed of light at the origin of coordinates where the gravitational potential is null, and $c$ is the frequency at the place where one observes a gravitational potential $\Phi$.

The experimental observation of both effects are complementary experimental proofs of the equivalence principle and of the equivalence between mass and energy. The Gravitational redshift has been observed in 1980 by Vessot et al. in the Gravity Probe-A (GP-A) experiment with an accuracy of $1.4\times 10^{-4}$ \cite{Vessot}, and the deflection of light by the Sun has been observed for the first time in 1919 by Eddington et al. \cite{Eddington}.

In the following we first evaluate the different ways in which gravitation affects the Casimir effect, to first and second order approximation. This sets the physics to devise two different experiments onboard a satellite in an high eccentric orbit, in order to experimentally probe the predicted theoretical influence of the gravitational redshift on Casimir cavities. We then discuss about the different theoretical aspects that these experiments can test and propose a dedicated space mission that we call Gravity-Probe C (GP-C).

\section*{On the Influence of Gravitation on the Casimir Force}
Usually one takes into account the Casimir effect only when looking for deviations of Newton's law of gravitation at micron and sub-$\mu m$ scales (cf. \cite{Fischbach} and references therein). However here we adopt a different approach: We propose to investigate the variation of the Casimir force as a result of a change of a gravitational potential experienced by the Casimir cavity.

The influence of gravitation on the Casimir effect was first studied by Sorge \cite{Sorge} and Calloni et al \cite{Calloni}.  Quach \cite{Quach} also recently studied the extension of the classical Casimir effect to gravitational waves.

To evaluate how the Casimir force is influenced by the gravitational redshift one can substitutes Equ.(\ref{e3}) directly into Equ.(\ref{e1}).
\begin{equation}
F=\frac{\pi^2}{240}\frac{\hbar}{a^4} c_0 \Big(1+\frac{\Phi}{c_0^2}\Big) A\label{e4}
\end{equation}
Setting $F_0=\frac{A\pi^2}{240}\frac{\hbar c_0}{a^4}$ as the Casimir force in the absence of gravitational field gradients, Equ.(\ref{e4}) simplifies to
\begin{equation}
\frac{\delta F}{F_0}=\frac{\Phi}{c_0^2}\label{e5}
\end{equation}
Where $\delta F=F-F_0$. One can also derive this relation by applying the gravitational redshift factor $\sqrt{g_{00}}=(1-2\Phi/c_0^2)^{1/2}$ to the energy of the different electromagnetic normal and transverse modes of the Casimir cavity. The total zero-point electromagnetic energy between two perfectly conducting plates can be written as
\begin{equation}
U=\hbar c_0 A \sum_{n} \int \frac {d^2k}{(2\pi)^2}\sqrt{(\sqrt{g_{00}}k)^2+\Big(\frac{n\pi\sqrt{g_{00}}}{a}\Big)^2}\label{e6}
\end{equation}
Where the normal modes are labeled with positive integers $n$ and the modes parallel to the conducting plates (transverse modes) are designated by their wave vector $k$. For $a<<r$ (where $r$ is the distance used to calculate the gravitational potential $\Phi$) one can make the approximation that between the plates of the Casimir cavity the red-shift factor is constant to first order: $\sqrt{g_{00}}\simeq(1+\Phi/c_0^2)$. This allows to reduce Equ.(\ref{e6}) to the form:
\begin{equation}
U=\Big(1+\frac{\Phi}{c_0^2}\Big) \hbar c_0 A \sum_{n} \int \frac {d^2k}{(2\pi)^2}\sqrt{k^2+\Big(\frac{n\pi}{a}\Big)^2}\label{e7}
\end{equation}
By calculating the integral by dimensional regularization, the energy takes the regular form of the Casimir expression (in Minkowski's spacetime) multiplied by the Gravitational redshift factor
\begin{equation}
U=\Big(1+\frac{\Phi}{c_0^2}\Big) \Big(-A \frac{\pi^2 \hbar c_0}{720 a^3}\Big) \label{e8}
\end{equation}
taking the minus of the gradient of $U$ with respect to $a$, $F=-\nabla_a U$, yields the red-shifted Casimir Force
\begin{equation}
F=\Big(1+\frac{\Phi}{c_0^2}\Big)\Big(A \frac{\pi^2 \hbar c_0}{240 a^4}\Big)=\Big(1+\frac{\Phi}{c_0^2}\Big)F_0 \label{e9}
\end{equation}
which is indeed identical to Equ.(\ref{e5}). One thus concludes that the attractive Casimir force between the conducting plates of a Casimir cavity should be different when the cavity is located in different gravitational potentials. This result is consistent with Sorge's first order approximation of Casimir pressure, if one takes duely into account (in Sorge's derivation) the effect of gravitation on the proper area of the Casimir plates \cite{Sorge} \footnote{Sorge demonstrated that the Casimir pressure on the cavity's plates (equivalent to the density of energy in the Casimir cavity) is not affected, to first order, by the gravitational redshift. This leads to a first order variation of the Casimir force according to Equ. (\ref{e9}), once we derive the Casimir force out of Sorge equation for Casimir pressure (Equ.(5.10) in reference \cite{Sorge})}.

It is to be noted that this effect is different from the one discussed by Calloni et al \cite {Calloni}. Calloni describes, to second order approximation, the effect of gravity on rigid Casimir cavities (with conducting plates constrained to remain at a constant separation distance), which are stationnary in the Earth gravitational field. He then calculates  a total net force which appears on a Casimir cavity as a result of the Earth gravitational redshift between its conducting plates. This force has a direction opposite to the Casimir cavity weight. It can be deduced assuming that the gravitational potential is not constant between the conducting plates of the cavity. Thus the two flat conducting plates forming the Casimir cavity will not experience the same Casimir attractive force. This results in a total Casimir force on the cavity which is different from zero and will be opposed to the cavity's weight.
\begin{equation}
\vec{F_{c}}=\Big(A \frac{\pi^2 \hbar c}{240 a^4}\Big)\frac{ag}{c^2} \vec{e}_r=F_0 \frac{\Delta \phi}{c^2} \label{e9a}
\end{equation}
where $g$ is the Earth gravitational acceleration, $\vec{e}_r$ is the unit vector in the $r$ direction, oriented with respect to the centre of the earth. Note that due to the smallness of the  separation distance between conducting plates, $a$, in a Casimir cavity, the gravitational potential difference in Equ.(\ref{e9a}) is in general several orders of magnitude below the gravitational potential difference in Equ.(\ref{e9}) for typical experimental figures. More generally being a second order effect, the Calloni net force much weaker than the deviation of the standard attractive Casimir force due to the gravitational redshift. This is precisely estimated in the following section.

\section*{Testing the Influence of Gravitation on the Casimir Effect with a Space Mission}
Is it possible to measure the dependence of the Casimir force on the gravitational potential to which the Casimir cavity is subjected? From Equ.(\ref{e5}) it is clear that the larger the gradient of gravitational potential $\Phi$ the larger should be the variations of the Casimir force on the conducting plates. Let us consider a Casimir cavity onboard a satellite orbiting the Earth in a High Elliptic Orbit (HEO). The largest variation of gravitational potential will occur between the perigee and the apogee of the orbit, to first order,
\begin{equation}
\frac{\Delta \Phi}{c_0^2}=\frac {GM}{c_0^2} \Big( \frac{1}{r_p}-\frac{1}{r_a}\Big) \label{m1}
\end{equation}
where $M=5.98\times 10^{24}$ kg is the Earth mass, $G$ is Newton's constant, $c_0$ is the speed of light  in vacuum (in the absence of gravitation), $r_p$ and $r_a$ are respectively the radius vectors of the perigee and apogee points of the satellite's orbit with respect to the center of mass of the Earth. Substituting Equ.(\ref{m1}) in Equ.(\ref{e5}) one obtains

\begin{equation}
\delta F = F-F_0=F_0\frac {GM}{c_0^2} \Big( \frac{1}{r_p}-\frac{1}{r_a}\Big) \label{m2}
\end{equation}

Typical values for HEO satellites  are $r_p=41600 Km$ and $r_a=7100 Km$ which corresponds to a gravitational redshift factor of $\Delta \Phi/c_0^2=5 \times 10^{-10}$. This is of course very small and outside of the measurement capabilities from present Casimir force measurement setups in the lab that approach accuracies down to 0.1\% for forces in the tens of pN between a flat surface and gold coated sphere on a cantilever \cite{Bordag2009}. However, a few possibilities can already be listed that may enable such measurements:

\begin{itemize}
\item Present measurements focus on the absolute accuracy to validate the Casimir effect for certain geometries and distances. However, we are looking for a relative accuracy as we want to investigate a change of the Casimir effect along the orbit. This may increase the sensitivity of the measurement.
\item Measurements are being done between flat surfaces and spheres as it is very difficult to align flat surfaces parallel to each other with sub-$\mu m$ distances. The environment of a drag-free satellite with ultraprecise attitude and orbit control such as LISA Pathfinder \cite{Antonucci2012} could make flat-flat plate measurements more practical as seismic noises and other common sources of error in the Earth laboratory are greatly reduced.
\item Recent studies propose the use of special metamaterials \cite{Zhao2011} or setups near power transmission lines \cite{Shahmoon2014} that are claimed to produce a giant Casimir effect with an amplification factor of up to $10^{11}$. If such material or setup is found, the measurement of our predicted effect can be done without problems.
\end{itemize}

We believe that such an experiment is very important as testing the influence of the gravitational redshift on the Casimir effect not only tests the laws of gravitation but also the laws of quantum mechanics, as well as alternative theories in both domains. Einstein \cite{Einstein} and Born \cite{Born} have shown that gravitational time dilation and speed of light variation in gravitational fields are intricate effects.

By investigating how the gravitational redshift affects the Casimir force, one is not only evaluating the equivalence principle for zero-point electromagnetic energy, but we are also pearing into the physical nature of mass. For example, as shown recently by one of us \cite{Tajmar}, Assis's gravity model leads to a  Planck constant which depends on the electron charge $e$, the speed of light in vacuum $c$ and the electric permittivity of the vacuum $\epsilon_0$.

\begin{equation}
\hbar = 3.62\frac{7 \pi^3 e^2}{72 c\epsilon_0}\label{m4}
\end{equation}

Substituting this expression for $\hbar$ into Equ.(\ref{e1}) one concludes that the Casimir effect should not depend on the value of the speed of light (with or without gravity)

\begin{equation}
F=1.47\times 10^{-3}\Big(\frac{\pi^4 e^2}{\epsilon_0 a^4} \Big)A\label{m5}
\end{equation}

In this case, to first order, the Casimir effect would not be at all affected by the gravitational redshift and it should not vary along the satellites orbit. Therefore if the experimentally measured variation of the Casimir force $\delta F$ in the space mission described above in Equ.(\ref{e9}) is null, thus showing no dependence on the gravitational potential and on the gravitational redshift, this could either become a supportive experimental fact in favor of a completely electromagnetic account of a fundamental quantum of mass, or reveal that zero-point electromagnetic energy violates the equivalence principle in terms of either Local Position Invariance(LPI) or Local Lorentz Invariance (LLI).

Interestingly, it is to be noted that a similar analysis to investigate the consequences of the Planck constant expression from Equ.(\ref{m4}) on Calloni's (second order) effect Equ.(\ref{e9a}) reveals that it will not change Calloni's prediction. Thus even if the standard Casimir force is unaffected by the gravitational redshift to first order a net Calloni's force on the Casimir Cavity should continue to appear. This indicates that the two different experiments discussed above are highly complementary in  order to derive consequences at theoretical level.

\section*{Conclusion: Gravity-Probe C}
We conclude that the influence of the gravitational redhsift on the Casimir effect is two fold: To first order approximation it modifies the usual Casimir attractive force between the conducting plates, and to second order it causes a differential net force on the overall cavity. Both effects could be observed onboard a satellite orbiting the Earth in a high eccentric orbit through the observation of the variation of the respective effects between the apogee and the perigee of the orbit. Substantial technology development is still necessary to reach the relative accuracy of $<5 \times 10^{-10}$ necessary to detect the effect. New concepts predicting a giant casimir effect may enable such technology.

We propose to designate the satellite hosting the two experiments presented above as Gravity Probe-Casimir, i.e. Gravity Probe-C or GP-C in short. In a nutshell Gravity Probe-C is a fundamental physics space mission that could deliver important experimental information about how gravitation physically influences the quantum vacuum, which is key to approach experimentally certain aspects of quantum gravitation.

\section*{Acknowledgement}
The authors would like to dedicate this paper to the celebration of 100 years of Einstein's theory of General Relativity.

\end{document}